\documentclass[12pt]{article}

\usepackage{amsmath,amssymb,amsbsy,amsfonts,latexsym,graphicx}
\usepackage{color,array}
\usepackage[nospace,sort,compress]{cite}
\usepackage{emptypage} 
\usepackage[usenames,dvipsnames]{xcolor}
\usepackage[unicode,psdextra]{hyperref}
\hypersetup{
    colorlinks=true, 
    citecolor=ForestGreen,
    linkcolor=RoyalBlue,
    urlcolor=RoyalBlue,
    pdfstartview=FitV,
    breaklinks=true,
    linktocpage=true}

\usepackage{bookmark}

\allowdisplaybreaks

\evensidemargin \oddsidemargin

\begin{document}


\begin{titlepage}

\renewcommand{\thefootnote}{\fnsymbol{footnote}}


\begin{flushright}
\end{flushright}

\vspace{15mm}
\baselineskip 9mm
\begin{center}
  {\Large \bf 1/2-BPS membrane instantons in AdS$_4 \times$
              S$^7 / \mathbf{Z}_k$}
\end{center}

\baselineskip 6mm
\vspace{10mm}
\begin{center}
Jaemo Park\footnote{\tt jaemo@postech.ac.kr} and
Hyeonjoon Shin\footnote{\tt nonchiral@gmail.com}
\\[10mm]
  {\sl Department of Physics \&
       Center for Theoretical Physics,\\
       POSTECH, Pohang, Gyeongbuk 37673, South Korea}
\\ and \\
  {\sl Asia Pacific Center for Theoretical Physics, \\
       Pohang, Gyeongbuk 37673, South Korea}
\end{center}

\thispagestyle{empty}

\vfill
\begin{center}
{\bf Abstract}
\end{center}
\noindent
According to the covariant open superstring description of D-branes in
the AdS$_4 \times \mathbf{CP}^3$ background, 1/2-BPS D2-branes are purely
instantonic.  Based on this and by taking the eleven dimensional
viewpoint, we identify the 1/2-BPS instantonic M2-brane configurations
in the AdS$_4 \times$ S$^7 / \mathbf{Z}_k$ background, which reduces to
the AdS$_4 \times \mathbf{CP}^3$ under the large $k$ limit, and evaluate
their action values.  We also consider the previously known 1/2-BPS
instantonic objects in ten dimensions from the M2-brane viewpoint
to compare with our results.
\\ [15mm]

\vspace{5mm}
\end{titlepage}

\baselineskip 6.6mm
\renewcommand{\thefootnote}{\arabic{footnote}}
\setcounter{footnote}{0}

\tableofcontents

\section{Introduction}

The covariant open superstring description of D-branes
\cite{Lambert:1999id,Bain:2002tq} is a useful tool
in classifying supersymmetric D-branes, especially 1/2-BPS D-branes,
in a given supersymmetric background.
It has been successfully applied to some important backgrounds
in superstring theories such as the flat spacetime \cite{Lambert:1999id},
IIB plane wave \cite{Bain:2002tq}, IIA plane wave \cite{Hyun:2002xe},
AdS$_5 \times$ S$^5$ \cite{Sakaguchi:2003py,Sakaguchi:2004md,
ChangYoung:2012gi,Hanazawa:2016lvo}, and AdS$_4 \times \mathbf{CP}^3$
\cite{Park:2018gop} backgrounds.
The data obtained after the classification of supersymmetric D-branes
are however `primitive' in a sense that they do not tell us about
which configuration of a given D-brane is really supersymmetric or
which part of the background supersymmetry is preserved on the D-brane
worldvolume.  Nevertheless, the classification provides us a good
guideline for further exploration of supersymmetric D-branes.  Indeed
this has been illustrated for the AdS$_5 \times$ S$^5$ background
in \cite{Park:2017ttx}.

In our previous work \cite{Park:2018gop}, we have obtained the data about
1/2-BPS D-branes in the AdS$_4 \times \mathbf{CP}^3$ background.
One interesting point from the data is that 1/2-BPS D2-brane is purely
instantonic and there is no 1/2-BPS Lorentzian D2-brane in contrast
to other D-branes of different dimensionalities.

As is well known, the Type IIA superstring theory in the
AdS$_4 \times \mathbf{CP}^3$ background is dual to the
Aharony-Bergman-Jafferis-Maldacena (ABJM) theory \cite{Aharony:2008ug}.
Under this correspondence, the study on the nonperturbative aspects of the
ABJM theory has been a fascinating research subject, in which some
instantonic objects have played important roles.\footnote{For comprehensive
review, see \cite{Marino:2016new}.}. Basically, there are two types of
instantonic objects called worldsheet and membrane instantons.  Both of
them are known to be 1/2-BPS and, in the bulk side, are identified as
string \cite{Cagnazzo:2009zh} and D2-brane \cite{Drukker:2011zy} instantons
wrapping certain subspaces of $\mathbf{CP}^3$.  By the way, the
result of \cite{Park:2018gop} indicates that, in addition to
these instantons, there is a possibility to have other 1/2-BPS instantonic
D2-brane configurations.

In the present work, we try to identify such
1/2-BPS D2-brane configurations purely based on the data obtained in
\cite{Park:2018gop} and evaluate their action values.  These 1/2-BPS D2
brane configurations can be lifted to membrane (M2-brane) configurations.
Then one can extrapolate 1/2 BPS M2-brane configurations
in the AdS$_4 \times$ S$^7 / \mathbf{Z}_k$ background for  all finite $k$, which reduces to
the AdS$_4 \times \mathbf{CP}^3$ in the large $k$ limit.
We also consider
the ten dimensional instantons of \cite{Cagnazzo:2009zh} and
\cite{Drukker:2011zy} and comment on the corresponding the M2-brane configurations.

In the next section, we give a brief description of the
AdS$_4 \times$ S$^7 / \mathbf{Z}_k$ background.  The possible 1/2-BPS
instantonic M2-brane configurations are identified in Sec.~\ref{bpsm2},
and their action values are evaluated in Sec.~\ref{instaction}.
Finally, the conclusion follows in Sec.~\ref{concl}.

\section{AdS$_4 \times$ S$^7 / \mathbf{Z}_k$ background}

The AdS$_4 \times$ S$^7 / \mathbf{Z}_k$ background originated from
the near horizon limit of M2-brane supergravity solution is composed of
the AdS$_4 \times$ S$^7 / \mathbf{Z}_k$ geometry
\begin{align}
ds^2 = \frac{R^2}{4} ds^2_{AdS_4} + R^2 ds^2_{S^7/\mathrm{Z}_k}
\end{align}
and the four-form field strength
\begin{align}
F_4 = \frac{3}{8} R^3 \epsilon_{AdS_4} \,,
\label{4form}
\end{align}
where $R$ is the radius of S$^7$ given by
\begin{align}
R = \ell_p (2^5 \pi^2 N k)^{1/6}
\label{s7rad}
\end{align}
with the eleven dimensional Planck length $\ell_p$ and the
number of M2-branes $Nk$, and $\epsilon_{AdS_4}$ is
the volume form of AdS$_4$ space.  The metric of AdS$_4$ is, in
the global coordinates,
\begin{align}
ds^2_{AdS_4} = - \cosh^2 \rho dt^2 + d \rho^2
              + \sinh^2 \rho ( d \theta^2 + \sin^2 \theta d \psi^2 ) \,.
\end{align}
As for the geometry of S$^7 / \mathbf{Z}_k$, , it is natural to express
it through the $U(1)$ Hopf fibration over $\mathbf{CP}^3$ for its obvious
connection with the Type IIA background \cite{Nilsson:1984bj}:
\begin{align}
\label{s7zk}
ds^2_{S^7/\mathrm{Z}_k} = ds^2_{\mathbf{CP}^3}
                         + \frac{1}{k^2} (dy + A)^2 \,,
\end{align}
where $y$ is the $U(1)$ fiber coordinate with the period
$y \sim y + 2\pi$, $A$ is a one-form given by
\begin{align}
\label{1form}
A = \frac{k}{2}
\left( \cos \alpha d \chi + \cos^2 \frac{\alpha}{2} \cos \theta_1 d\phi_1
+ \sin^2 \frac{\alpha}{2} \cos \theta_2 d\phi_2 \right) \,,
\end{align}
and the $\mathbf{CP}^3$ geometry is as follows:
\begin{align}
ds^2_{\mathbf{CP}^3} =
& \frac{1}{4} d\alpha^2
  + \frac{1}{4} \cos^2 \frac{\alpha}{2}
         (d\theta_1^2 + \sin^2 \theta_1 d\phi_1^2)
  + \frac{1}{4} \sin^2 \frac{\alpha}{2}
         (d\theta_2^2 + \sin^2 \theta_2 d\phi_2^2)
\notag \\
& + \frac{1}{4} \sin^2 \frac{\alpha}{2} \cos^2 \frac{\alpha}{2}
         (2 d \chi + \cos \theta_1 d \phi_1 - \cos \theta_2 d \phi_2)^2 \,,
\label{cp3}
\end{align}
with the ranges of angles, $0 \le \alpha, \theta_1, \theta_2 \le \pi$
and $0 \le \chi, \phi_1, \phi_2 \le 2\pi$.

From the metric of (\ref{s7zk}),
we choose the elfbeine for S$^7 / \mathbf{Z}_k$
as\footnote{$e^{0,1,2,3}$
are for the AdS$_4$ space.  Although their explicit form is not necessary
in this work, if we follow the choice of \cite{Drukker:2011zy}, they are
$e^0 = \frac{1}{2} \cosh \rho dt$, $e^1 = \frac{1}{2} d\rho$,
$e^2 = \frac{1}{2} \sinh \rho d \theta$, and
$e^3 = \frac{1}{2} \sinh \rho \sin \theta d \psi$.}
\begin{align}
& e^4 = \frac{1}{2} d \alpha \,, \quad
e^5 = \frac{1}{2} \sin \frac{\alpha}{2} \cos \frac{\alpha}{2}
         ( 2 d \chi + \cos \theta_1 d \phi_1 - \cos \theta_2 d \phi_2) \,,
\notag \\
& e^6 = \frac{1}{2} \cos \frac{\alpha}{2} d \theta_1 \,, \quad
e^7 = \frac{1}{2} \cos \frac{\alpha}{2} \sin \theta_1 d \phi_1 \,,
\notag \\
& e^8 = \frac{1}{2} \sin \frac{\alpha}{2} d \theta_2 \,, \quad
e^9 = \frac{1}{2} \sin \frac{\alpha}{2} \sin \theta_2 d \phi_2 \,,
\notag \\
& e^{\natural} = -\frac{1}{k} (dy + A) \,.
\end{align}
The reason for this choice is to make the K\"{a}hler structure of
$\mathbf{CP}^3$ in a canonical form.  The K\"{a}hler structure itself can
be read off from one property of one-form $A$, $dA = 2 k J$,
where $J$ is the K\"{a}hler two-form of $\mathbf{CP}^3$.  If we compute
$dA$ using (\ref{1form}), we obtain
\begin{align}
dA = -2k \left( e^4 \wedge e^5 + e^6 \wedge e^7 + e^8 \wedge e^9 \right) \,,
\label{kahler}
\end{align}
which indeed shows that $J$ has the block diagonal structure, that is,
the canonical form and thus the above choice for the elfbeine is a
proper one.

\section{1/2-BPS M2-brane instantons}
\label{bpsm2}

For the investigation of 1/2-BPS M2-brane configurations, we need
the Killing spinor $\epsilon$ of the AdS$_4 \times$ S$^7 / \mathbf{Z}_k$
background.  It has been obtained
in \cite{Nishioka:2008ib,Drukker:2008zx,Drukker:2011zy}
by solving the Killing spinor equation and has the following form.
\begin{align}
\epsilon = \mathcal{M} \epsilon_0 \,,
\label{kspinor}
\end{align}
where $\epsilon_0$ is a 32 component constant
spinor\footnote{The constant spinor $\epsilon_0$ has 24 free components
for $k > 2$. Otherwise, that is, for $k=1,2$, it has 32 free components.
For more detailed discussion, see Refs.
\cite{Nishioka:2008ib,Drukker:2008zx,Drukker:2011zy}.}
and $\mathcal{M}$ is given by
\begin{align}
\mathcal{M} =
& e^{ \frac{\alpha}{4} (\hat{\gamma} \gamma_4 - \gamma_{5\natural}) }
  e^{ \frac{\theta_1}{4} (\hat{\gamma} \gamma_6 - \gamma_{7\natural}) }
  e^{ \frac{\theta_2}{4} (\gamma_{59} + \gamma_{48}) }
  e^{ -\frac{\xi_1}{2} \hat{\gamma} \gamma_\natural }
  e^{ -\frac{\xi_2}{2} \gamma_{67} }
  e^{ -\frac{\xi_3}{2} \gamma_{45} }
  e^{ -\frac{\xi_4}{2} \gamma_{89} }
\notag \\
& \times
  e^{ \frac{\rho}{2} \hat{\gamma} \gamma_1 }
  e^{ \frac{t}{2} \hat{\gamma} \gamma_0 }
  e^{ \frac{\theta}{2} \gamma_{12} }
  e^{ \frac{\psi}{2} \gamma_{23} }
\label{km}
\end{align}
with $\hat{\gamma}=\gamma^{0123}$ and a set of coordinate combinations,
\begin{align}
\xi_1 = \frac{1}{2} ( \phi_1 + \chi + 2y) \,, \quad
\xi_2 = \frac{1}{2} ( -\phi_1 + \chi + 2y) \,,
\notag \\
\xi_3 = \frac{1}{2} ( \phi_2 - \chi + 2y) \,, \quad
\xi_4 = \frac{1}{2} ( -\phi_2 - \chi + 2y) \,.
\end{align}
The Killing spinor $\epsilon$ is obtained in the Lorentzian
signature.  Since we are concerned about the instantonic configuration
for which the spacetime is taken to be Euclidean, we should recast
$\epsilon$ in a way to respect the Euclidean nature.  However, we
will not care much about it because the time coordinate will be set
to zero and our interest is the consistent projection operators acting
on $\epsilon$ (strictly speaking $\epsilon_0$) which identify the
1/2-BPS configurations.

Having the Killing spinor, the 1/2-BPS instantonic M2-brane configurations
can be considered by using the usual equation
\begin{align}
\Gamma \epsilon = \epsilon \,,
\label{half}
\end{align}
which is obtained by combining the spacetime supersymmetry and
$\kappa$-symmetry transformation.
The symbol $\Gamma$ is the spinorial matrix appearing in the
$\kappa$-symmetry projector and satisfies $\Gamma^2=1$ and
$\mathrm{Tr}\Gamma=0$.
The explicit expression of $\Gamma$ for M2-brane is
\begin{align}
\Gamma = \frac{i}{3! \sqrt{g}} \epsilon^{ijk}
         \Pi^a_i \Pi^b_j \Pi^c_k \Gamma_{abc}
\label{gexp}
\end{align}
where $\Pi^a_i = \partial_i X^\mu e^a_\mu$ and $g$ is the
determinant of the induced metric on M2-brane,
$g_{ij} = \Pi^a_i \Pi^b_j \eta_{ab}$.  The indices $i,j,k$ are those of
the M2-brane worldvolume, and $\mu,\nu,\dots$ ($a,b,\dots$) are
the curved (tangential) spacetime indices.
Now, by using the Killing spinor (\ref{kspinor}), the Eq.~(\ref{half})
can be rewritten as
\begin{align}
\tilde{\Gamma} \epsilon_0 = \epsilon_0 \,,
\label{half0}
\end{align}
where we have defined
\begin{align}
\tilde{\Gamma} \equiv \mathcal{M}^{-1} \Gamma \mathcal{M} \,.
\label{gtilde}
\end{align}
Here $\tilde{\Gamma}^2 = 1$ is guaranteed because $\Gamma^2 =1$.
For a given M2-brane configuration, $\Gamma$ and $\mathcal{M}$ have
the corresponding expressions by which $\tilde{\Gamma}$ is
determined. If the resulting $\tilde{\Gamma}$ does not depend on any
worldvolume coordinate, then the M2-brane configuration
is confirmed to be 1/2-BPS.

For specifying a M2-brane configuration, let us introduce a notation
\begin{align}
[ X, Y, Z ] \,,
\end{align}
which means that the M2 worldvolume coordinates
($\zeta^1$, $\zeta^2$, $\zeta^3$) are identified as
$\zeta^1 = X$, $\zeta^2 = Y$, and
$\zeta^3 = Z$. In other words, the notation represents a static
gauge for the worldvolume reparametrization.  Except for the coordinates
along which M2-brane spans, all other coordinates in $\mathcal{M}$ which
are transverse to M2-brane are set to zero. These are the natural
generalizations of the previous results of 1/2-BPS D2-brane configurations,
which can be easily verified in M theory setting.
However, if a polar coordinate among $\alpha$, $\theta_1$,
$\theta_2$ is taken to be transverse one,
it is kept to be an arbitrary constant.

Before investigating the instantonic M2-brane configurations
based on the covariant open string description of 1/2-BPS D-branes,
we briefly reconsider the previously explored string worldsheet
instanton \cite{Cagnazzo:2009zh} and D2-brane
instanton \cite{Drukker:2011zy} from the viewpoint of eleven dimensional
M2-brane.  Both of them have been studied in the
context of ten dimensional IIA string theory and turned out to be 1/2-BPS.
As for the string worldsheet instanton, it wraps $\mathbf{CP}^1$ ($\subset$
$\mathbf{CP}^3$) parametrized by $\alpha$ and $\chi$.  Since a string
is nothing but the M2-brane wrapping the M-theory circle direction $y$,
the corresponding M2-brane configuration is given by
$[\alpha, \chi, y]$.  The expression of $\Gamma$ (\ref{gexp})
becomes simply $\Gamma = -i \gamma_{45\natural}$ and $\tilde{\Gamma}$ of
Eq.~(\ref{gtilde}) is evaluated as
\begin{align}
\tilde{\Gamma} = -i \gamma_{45\natural} \,.
\label{wsg}
\end{align}
By the way, since there are two more two spheres within
$\mathbf{CP}^3$ parametrized by $(\theta_1, \phi_1)$ and
$(\theta_2, \phi_2)$, one may be curious about whether the configurations
$[\theta_1, \phi_1, y]$ and $[\theta_2, \phi_2, y]$ are also 1/2-BPS.
Actually, they are but in a restricted sense.  As for
$[\theta_1, \phi_1, y]$, it is not difficult to check that the
configuration is 1/2-BPS only when its transverse position in $\alpha$ is
zero, $\alpha=0$.\footnote{One can check that the resulting two
sphere has maximal radius $R/2$ when  $\alpha=0$.}
 In this case, $\Gamma = -i \gamma_{67\natural}$ and
$\tilde{\Gamma}$ simply becomes $-i \gamma_{67\natural}$.
As for another configuration $[\theta_2, \phi_2, y]$, it is 1/2-BPS
only when $\alpha = \pi$. For this, we get
$\Gamma = -i \gamma_{89\natural}$ and the corresponding $\tilde{\Gamma}$
as $i \hat{\gamma} \gamma_{4589} = -i \gamma_{67\natural}$.\footnote{An
identity $\hat{\gamma}\gamma_{456789} = \gamma_\natural$ has been
used.}

On the other hand, the D2-brane instanton spans the Lagrangian
submanifold $\mathbf{RP}^3$ within $\mathbf{CP}^3$ which
is parametrized by the coordinates ($\chi$, $\vartheta$, $\varphi$)
with the identifications $\vartheta = \theta_1 = \theta_2$ and
$\varphi = \phi_1 = -\phi_2$.  Since a D2-brane in ten dimensions
is just an M2-brane in eleven dimensions, the M2-brane configuration
corresponding to the D2-brane instanton is simply
$[ \chi, \vartheta, \varphi ]$.
Then the $\Gamma$ of (\ref{gexp}) for this configuration becomes
$\Gamma = - i \gamma_{67 \natural}
 e^{ \frac{\alpha}{2} ( 2 \gamma_{5 \natural} + \gamma_{68} - \gamma_{79})}$
and the evaluation of $\tilde{\Gamma}$ of (\ref{gtilde}) leads to
\begin{align}
\tilde{\Gamma} =
- i \gamma_{67 \natural}
e^{ \frac{\alpha}{2} ( \hat{\gamma} \gamma_4 + \gamma_{5 \natural}
    + \gamma_{68} - \gamma_{79} )} \,.
\label{d2g}
\end{align}
Obviously, the above two $\tilde{\Gamma}$'s, (\ref{wsg}) and (\ref{d2g}),
are independent of any of the worldvolume coordinates.  Thus, from
Eq.~(\ref{half0}), it is confirmed that the previously known M2-brane
configurations are 1/2-BPS as they should be.

Now let us turn to our main concern and investigate the 1/2-BPS
M2-brane instantonic configurations based on the covariant open string
description of D-branes.  According to the open string description
\cite{Park:2018gop}, purely instantonic D2-brane can be 1/2-BPS
under the following condition: only one worldvolume direction is allowed
to extend in each of three two dimensional subspaces of $\mathbf{CP}^3$.
The three subspaces are realized by looking at the K\"{a}hler structure
of $\mathbf{CP}^3$ (\ref{kahler}).  Following this condition,
we see that there are eight candidates for the 1/2-BPS M2-brane
instantonic configurations which are
$[\alpha, \theta_1, \theta_2]$, $[\alpha, \theta_1, \phi_2]$,
$[\alpha, \phi_1, \theta_2]$, $[\alpha, \phi_1, \phi_2]$,
$[\chi, \theta_1, \theta_2]$, $[\chi, \theta_1, \phi_2]$,
$[\chi, \phi_1, \theta_2]$, and $[\chi, \phi_1, \phi_2]$.

We have investigated all the candidates and found that only four
configurations in which $\alpha$ is a worldvolume direction are 1/2-BPS.
In what follows, we describe the 1/2-BPS configurations in sequence.
First of all, for the configuration $[ \alpha, \theta_1, \theta_2 ]$,
the $\Gamma$ of (\ref{gexp}) becomes $\Gamma = i \gamma_{468}$, and
the corresponding $\tilde{\Gamma}$ of (\ref{gtilde}) is evaluated as
\begin{align}
\tilde{\Gamma} = i \gamma_{468} \,.
\label{m21}
\end{align}
Second, the configuration $[ \alpha, \theta_1, \phi_2 ]$ with fixed
$\theta_2$ leads to
$\Gamma = i \gamma_{456} e^{-\frac{\alpha}{4} \gamma_{5 \natural}}
e^{-\theta_2 \gamma_{59}} e^{-\frac{\alpha}{4} \gamma_{5 \natural}}$,
and we obtain the corresponding $\tilde{\Gamma}$ as
\begin{align}
\tilde{\Gamma} = i \gamma_{456} e^{\frac{\theta_2}{2}
(\gamma_{48} - \gamma_{59}) } \,.
\label{m22}
\end{align}
The third configuration $[ \alpha, \phi_1, \theta_2 ]$ is equivalent
to $[ \alpha, \theta_1, \phi_2 ]$.  So we just move on to the last
configuration $[ \alpha, \phi_1, \phi_2 ]$ with fixed $\theta_1$ and
$\theta_2$.  For this configuration, we have
$\Gamma = - i \gamma_{45 \natural}
     e^{- \frac{\alpha}{4} \gamma_{5 \natural}}
     e^{\theta_1 \gamma_{7 \natural}}
     e^{- \theta_2 \gamma_{59}}
     e^{\frac{\alpha}{4} \gamma_{5 \natural}}$,
and the corresponding $\tilde{\Gamma}$ is obtained as
\begin{align}
\tilde{\Gamma} =
   - i \gamma_{45 \natural}
     e^{\frac{\theta_1}{2} ( \hat{\gamma} \gamma_6 + \gamma_{7 \natural} )}
     e^{\frac{\theta_2}{2} ( \gamma_{48} - \gamma_{59} ) } \,.
\label{m23}
\end{align}
We see that all the resulting $\tilde{\Gamma}$'s in (\ref{m21}),
(\ref{m22}), and (\ref{m23}) do not depend on any of the worldvolume
coordinates.  Therefore Eq.~(\ref{half0}) shows that half of the
spacetime supersymmetry $(1 + \tilde{\Gamma}) \epsilon_0$ is preserved
on the M2-brane worldvolume.

At this point, one may ask why the other four configurations in which
$\chi$ is a worldvolume direction are not 1/2-BPS. Although we have
checked all of them, we just take one of them as an example and
briefly illustrate the reason. Consider
$[ \chi, \theta_1, \theta_2 ]$.  For this
configuration, $\Gamma$ of (\ref{gexp}) is simply
$\Gamma = - i \gamma_{68 \natural}$
and $\mathcal{M}$ of (\ref{km}) becomes
$\mathcal{M} =
e^{\frac{\theta_1}{4} (\hat{\gamma} \gamma_6 - \gamma_{7 \natural})}
e^{\frac{\theta_2}{4} (\gamma_{59} + \gamma_{48} )}
e^{ - \frac{\chi}{4} (\hat{\gamma} \gamma_\natural + \gamma_{67}
                      - \gamma_{45} - \gamma_{89} ) }
$.
The evaluation of $\tilde{\Gamma}$ of (\ref{gtilde}) is the process of
pushing $\Gamma$ to the left of $\mathcal{M}^{-1}$ (or to the right of
$\mathcal{M}$).  But, unlike the previous cases, this process makes
us face with a problem from the beginning.  That is, what we get
for the $\theta_1$ dependent part is
\begin{align}
e^{- \frac{\theta_1}{4} (\hat{\gamma} \gamma_6 - \gamma_{7 \natural})}
\gamma_{68 \natural}
e^{\frac{\theta_1}{4} (\hat{\gamma} \gamma_6 - \gamma_{7 \natural})}
= \gamma_{68 \natural}
e^{- \frac{\theta_1}{2} \gamma_{7 \natural}} \,,
\end{align}
which means that $\tilde{\Gamma}$ depends on the worldvolume coordinate
$\theta_1$.  There is no way to eliminate the $\theta_1$ dependence.
This is also the case for other worldvolume coordinates, $\theta_2$
and $\chi$.  Thus $\tilde{\Gamma}$ depends on the worldvolume
coordinates implying that we have different projection operators
at different worldvolume positions.  This inconsistency explicitly shows
that the M2-brane configuration $[ \chi, \theta_1, \theta_2 ]$ is
not 1/2-BPS.  Of course, the configuration may be less supersymmetric
by imposing other projection conditions.  However, since here we are
interested in 1/2-BPS objects, it will
not be considered further.

\section{Instanton action}
\label{instaction}

For 1/2-BPS M2-brane instanton configurations considered in the last
section, let us evaluate their classical action values.  The basic
purpose is to compare the resulting values with those for the string
worldsheet instanton and D2-brane instanton obtained in
\cite{Cagnazzo:2009zh} and \cite{Drukker:2011zy} respectively.

The bosonic part of the Euclidean M2-brane action is given by
\begin{align}
S = T_2 \int_{\mathcal{M}_3} d^3 \zeta \sqrt{g}
   + i T_2
     \int_{\mathcal{M}_4 \, (\mathcal{M}_3 = \partial \mathcal{M}_4)}
          F_4 \,,
\label{m2-action}
\end{align}
where $T_2$ is the M2-brane tension,
\begin{align}
T_2 = \frac{1}{4 \pi^2 \ell_p^3} \,,
\label{m2tension}
\end{align}
and $g$ is the determinant of the induced metric on the M2-brane
worldvolume.  Since the M2-brane instantons we have considered are placed
within S$^7 / \mathbf{Z}_k$ while the background four-form field
strength $F_4$ is turned on in the AdS$_4$ space as one can see from
Eq.~(\ref{4form}), the Wess-Zumino term including $F_4$ does not contribute
to the action.  Thus we only need to evaluate the Nambu-Goto type kinetic
term.

We first consider the configuration $[ \alpha, \phi_1, \phi_2 ]$ with
fixed $\theta_1$ and $\theta_2$.  For this, $\sqrt{g}$ is computed as
$\frac{R^3}{8} \cos \frac{\alpha}{2} \sin \frac{\alpha}{2}$. We see that
there is no dependence on the transverse coordinates $\theta_1$ and
$\theta_2$, and hence the action value will always be the same
regardless of the transverse position of the M2-brane configuration.
Having the expression for $\sqrt{g}$, it is straightforward to evaluate
the action, which is obtained as follows:
\begin{align}
S &= 2 T_2 \int^\pi_0 d \alpha \int^{2\pi}_0 d \phi_1 \int^{2\pi}_0 d \phi_2
    \frac{R^3}{8} \cos \frac{\alpha}{2} \sin \frac{\alpha}{2}
\notag \\
  & = \pi^2 T_2 R^3
\notag \\
  & = \pi k \sqrt{ \frac{2N}{k} } \,,
\label{m2value}
\end{align}
where Eqs.~(\ref{s7rad}) and (\ref{m2tension}) have been used
in the last step.  We note that the $\alpha$ integration has
doubled in the first line as
$\int^\pi_0 d\alpha \rightarrow 2 \int^\pi_0 d\alpha$
basically because the nature of $\alpha$ is a polar angle.
The two angles $\alpha$ and $\chi$ parametrize a sphere as one
notices from Eq.~(\ref{cp3}).  One worldvolume direction of the present
M2-brane wraps this sphere along $\alpha$ with fixed $\chi$.
But, since $\alpha$ is a polar angle, the worldvolume direction covers
not a circle but half the circle geometrically.  Thus, in order to get
the correct action value, we should double the integration over $\alpha$.

By following the same way, the action values for other configurations
$[\alpha, \theta_1, \theta_2]$, $[\alpha, \theta_1, \phi_2]$, and
$[\alpha, \phi_1, \theta_2]$ can be evaluated.  The resulting values
are exactly the same with that of $[ \alpha, \phi_1, \phi_2 ]$
obtained in Eq.~(\ref{m2value}).  We note that the action value
$\pi k \sqrt{2N/k}$ agrees with that for the D2-brane instanton wrapping
the Lagrangian submanifold $\mathbf{RP}^3$ within $\mathbf{CP}^3$
computed in \cite{Drukker:2011zy}.  Although the D2-brane result
of \cite{Drukker:2011zy} has been obtained in ten dimensional
IIA theory, it is not difficult to check that it holds even in eleven
dimensions without any modification by considering the M2-brane
configuration corresponding to the D2-brane wrapping the $\mathbf{RP}^3$.
Thus it seems plausible that all the 1/2-BPS M2-brane instantons which
become D2-brane instantons in ten dimensions are
characterized by the action value $\pi k \sqrt{2N/k}$.

A noteworthy point of the above result is that the action value is
valid even at finite $k$ in contrast to the ten dimensional case
where $k \gg 1$ is assumed for the validity of Type IIA
AdS$_4 \times \mathbf{CP}^3$ background.
Actually, this situation continues to hold even for the case of worldsheet instanton for which the action value has been obtained as
\cite{Cagnazzo:2009zh}
\begin{align}
S = 2 \pi \sqrt{ \frac{2N}{k} }  \,.
\end{align}
If we evaluate the corresponding M2-brane configuration for the
worldsheet instanton, that is, $[ \alpha, \chi, y]$, then we get the same
result $2 \pi \sqrt{2N/k}$ yet valid at finite $k$.\footnote{We
note that the configurations $[\theta_1, \phi_1, y]$ at $\alpha=0$
and $[\theta_2, \phi_2, y]$ at $\alpha = \pi$ considered in the
last section also have the same action value.}
Thus we see that the action values of the ten dimensional 1/2-BPS
instantons are not modified at finite $k$.

\section{Conclusion}
\label{concl}

Based on the data \cite{Park:2018gop} obtained from the covariant open
superstring description of 1/2-BPS D-branes in Type IIA
AdS$_4 \times \mathbf{CP}^3$ background,
we have explored the possible 1/2-BPS M2-brane instanton configurations
from the eleven dimensional viewpoint. It has been shown that there
exist four additional 1/2-BPS M2-brane instanton configurations in
addition to the previously known ones.  All of them, which are interpreted
as D2-brane instantons in ten dimensions, have the same action
value $\pi k \sqrt{2N/k}$ which is identical to that of D2-brane instanton
studied in \cite{Drukker:2011zy}.  One important result is that the
action value is valid even at finite $k$.  Actually, this is also the case
for the 1/2-BPS M2-brane instanton configuration corresponding to the
worldsheet instanton \cite{Cagnazzo:2009zh}.  We speculate that the validity
of action values at finite $k$ is due to large amount
of supersymmetry.

A supersymmetric brane configuration means that the theory on its woldvolume
is supersymmetric.  This in turn implies that there is no tachyonic mode on the
worldvolume signaling instability of the configuration.  Here, since we have
considered the Euclidean theory, it would be appropriate to view a
supersymmetric configuration as a minimal action configuration rather
than stable one.  Let us now read off the quadratic parts of small bosonic 
transverse deformations for each supersymmetric configuration based on the 
M2-brane action (\ref{m2-action}).  What we obtain is that all the transverse modes
are massless for all the supersymmetric configurations and their actions are of the
form $\int d^3 \zeta \sqrt{h} h^{ij} \partial_i \delta \varphi \partial_j \delta \varphi$
where $h^{ij}$ is the induced worldvolume metric for a given supersymmetric
configuration and $\delta \varphi$ represents the small transverse deformation.
Since the quadratic action is positive definite, the transverse deformations
tend to increase the action value.  This shows that the supersymmetric
instantons explored in this work are action minimizing ones. It has been known that 
for the supergravity background with nontrivial fluxes, the supersymmetric cycle
satisfies the generalized calibration condition, which satisfies action minimizing 
condition. Under the generalized calibration condition, it has been shown that one 
can have stable and supersymmetric cycles even though cycles are not carrying any 
topological charges \cite{Uranga}. Note that three cycles on $\mathbf{CP}^3$ do 
not carry integer charge since the third integer homology
is at most torsion. However  since the instantonic configurations we have considered 
are supersymmetric and therefore satisfy the generalized calibration condition, 
they are nevertheless favored ones.

\section*{Acknowledgments}


This research was supported by Basic Science Research Program through the
National Research Foundation of Korea (NRF) funded by the Ministry of
Education with Grant No.~NRF-2015R1A2A2A01007058, No.~2018R1A2B6007159 (JP),
and No.~2018R1D1A1B07045425 (HS).

%



\providecommand{\href}[2]{#2}\begingroup\raggedright
\endgroup

\end{document}